\documentclass[dvips,12pt,a4paper]{article}
\usepackage{a4wide}
\usepackage{epsfig}
\usepackage{amsmath}
\usepackage{latexsym}

  \newlength{\absize}
\newcommand{\dd}{\mbox{{\rm d}}}

\newcommand{\Lumint}{{\cal L}_{\rm int}}

\allowdisplaybreaks 

\catcode`@=11
\def\citer{\@ifnextchar [{\@tempswatrue\@citexr}{\@tempswafalse\@citexr[]}}
 
\def\@citexr[#1]#2{\if@filesw\immediate\write\@auxout{\string\citation{#2}}\fi
  \def\@citea{}\@cite{\@for\@citeb:=#2\do
    {\@citea\def\@citea{--\penalty\@m}\@ifundefined
       {b@\@citeb}{{\bf ?}\@warning
       {Citation `\@citeb' on page \thepage \space undefined}}%
\hbox{\csname b@\@citeb\endcsname}}}{#1}}
\catcode`@=12

\begin{document}
  \thispagestyle{empty}
  \pagestyle{empty}
  \renewcommand{\thefootnote}{\fnsymbol{footnote}}
\newpage\normalsize
    \pagestyle{plain}
    \setlength{\baselineskip}{4ex}\par
    \setcounter{footnote}{0}
    \renewcommand{\thefootnote}{\arabic{footnote}}
\newcommand{\preprint}[1]{%
  \begin{flushright}
    \setlength{\baselineskip}{3ex} #1
  \end{flushright}}
\renewcommand{\title}[1]{%
  \begin{center}
    \LARGE #1
  \end{center}\par}
\renewcommand{\author}[1]{%
  \vspace{2ex}
  {\Large
   \begin{center}
     \setlength{\baselineskip}{3ex} #1 \par
   \end{center}}}
\renewcommand{\thanks}[1]{\footnote{#1}}
\renewcommand{\abstract}[1]{%
  \vspace{2ex}
  \normalsize
  \begin{center}
    \centerline{\bf Abstract}\par
    \vspace{2ex}
    \parbox{\absize}{#1\setlength{\baselineskip}{2.5ex}\par}
  \end{center}}

\begin{flushright}
{\setlength{\baselineskip}{2ex}\par
{hep-ph/0107159} \\[2mm]
{July 2001}           \\
} 
\end{flushright}
\vspace*{4mm}
\vfill
\title{New physics signatures at a Linear Collider: model-independent 
analysis from `conventional' polarized observables}
\vfill
\author{
A.A. Babich$^{a}$,
P. Osland$^{b}$,
A.A. Pankov$^{a,c,d}$ {\rm and}
N. Paver$^{c}$}
\begin{center}
$^a$ Pavel Sukhoi Technical University, 
     Gomel, 246746 Belarus \\
$^b$ Department of Physics, University of Bergen, \\
     All\'{e}gaten 55, N-5007 Bergen, Norway \\
$^c$ Dipartimento di Fisica Teorica, Universit\`a di Trieste and \\
Istituto Nazionale di Fisica Nucleare, Sezione di Trieste,
Trieste, Italy\\
$^d$ The Abdus Salam International Centre for Theoretical Physics,  Trieste,
Italy
\end{center}
\vfill
\abstract
{We discuss four-fermion contact-interaction searches in the processes  
$e^+e^-\to\mu^+\mu^-$, $c{\bar c}$ and $b{\bar b}$ at a future 
$e^+e^-$ Linear Collider with c.m.\ energy $\sqrt{s}=0.5$ TeV and 
with both beams longitudinally polarized. Our analysis is based on the 
measurements of familiar polarized observables such as the total cross 
section and the forward-backward/left-right asymmetries, and accounts 
for the general set of contact interaction couplings as independent, 
non-zero, parameters thus avoiding simplifying, model-dependent, 
assumptions. We derive the corresponding model-independent constraints 
on the above-mentioned coupling constants, and evaluate the corresponding 
reach at the Linear Collider, emphasizing the role of beam polarization. 
We compare the results with a model-dependent procedure where only one 
coupling is varied at a time.
 }
\vspace*{20mm}
\setcounter{footnote}{0}
\vfill

\newpage
    \setcounter{footnote}{0}
    \renewcommand{\thefootnote}{\arabic{footnote}}
    \setcounter{page}{1}

\section{Introduction}
Contact interaction Lagrangians (CI) generally represent an effective 
description of the `low energy' manifestations of some non-standard 
dynamics acting at new, intrinsic, mass scales much higher than the 
energies reachable at current particle accelerators. As such, they can 
be studied through deviations of the experimental observables from the 
Standard Model (SM) expectation, that reflect the additional presence of 
the above-mentioned new interaction. 
Typical examples are the composite models and the exchanges of extremely 
heavy neutral gauge bosons and 
leptoquarks \cite{Barger:1998nf,Altarelli:1997ce}.    
\par 
Clearly, such deviations are expected to be extremely small, as they 
would be suppressed for dimensional reasons by essentially some power of
the ratio between the available energy and the large mass scales. 
Accordingly, very high energy reactions in experiments with high 
luminosity are one of the natural tools to investigate signatures of 
contact interaction couplings.  
In general, these constants are considered as {\it a priori} free 
parameters, and one can quantitatively derive an assessment of the attainable 
reach and of the corresponding upper limits, essentially by numerically 
comparing the deviations with the expected experimental statistical and 
systematical uncertainties on the cross sections.   
\par 
Here, we consider the fermion pair production process  
\begin{equation}
e^++e^-\to f+\bar{f}, \label{proc}
\end{equation}
with $f\ne e$, $t$, at a Linear Collider (LC) with c.m.\ energy 
$\sqrt s=0.5\hskip 2pt{\rm TeV}$ and polarized electron and positron 
beams. We discuss the sensitivity of this reaction to the general, 
$SU(3)\times SU(2)\times U(1)$ 
symmetric $eeff$ contact-interaction effective Lagrangian, with 
helicity-conserving and flavor-diagonal fermion 
currents \cite{Eichten:1983hw}:       
\begin{equation}
{\cal L}_{\rm CI}
=\sum_{\alpha\beta}g^2_{\rm eff}\hskip 2pt\epsilon_{\alpha\beta}
\left(\bar e_{\alpha}\gamma_\mu e_{\alpha}\right)
\left(\bar f_{\beta}\gamma^\mu f_{\beta}\right).
\label{lagra}
\end{equation}
In Eq.~(\ref{lagra}): $\alpha,\beta={\rm L,R}$ denote left- or 
right-handed helicities, generation and color indices have been suppressed,
and the CI coupling constants are parameterized in terms of corresponding 
mass scales as 
$\epsilon_{\alpha\beta}={\eta_{\alpha\beta}}/{{\Lambda^2_{\alpha\beta}}}$
with $\eta_{\alpha\beta}=\pm 1,0$ depending on the chiral structure of the
individual interactions. Also, conventionally the value of $g^2_{\rm eff}$ 
is fixed at $g^2_{\rm eff}=4\pi$, as a reminder that, in the case of
compositeness, the new interaction would become strong at $\sqrt s$ of the 
order of $\Lambda_{\alpha\beta}$. Obviously, in this parameterization, 
exclusion ranges or upper limits on the CI couplings can be equivalently
expressed as exclusion ranges or lower bounds on the corresponding mass 
scales $\Lambda_{\alpha\beta}$.   
\par 
For a given final fermion flavor $f$, ${\cal L}_{\rm CI}$ in 
Eq.~(\ref{lagra}) envisages the existence of eight individual, 
and independent, CI models corresponding to the combinations of the four 
chiralities $\alpha,\beta$ with the $\pm$ signs of the $\eta$'s,  
with {a priori} free, and nonvanishing, coefficients. Correspondingly, 
the most general (and model-independent) analysis of the process 
(\ref{proc}) must account for the complicated situation where all 
four-fermion effective couplings defined in Eq.~(\ref{lagra}) are 
simultaneously allowed in the expression for the cross section, and in 
principle can interfere and weaken the bounds in case of accidental 
cancellations. 
\par 
Of course, the different helicity amplitudes, as such, do not interfere.
However, {\it the deviations from the SM} could be positive for one
helicity amplitude, and negative for another.
Thus, cancellations might occur.
\par 
The simplest attitude is to assume non-zero values for only one of the 
couplings (or one specific combination of them) at a time, with all others 
zero, this leads to tests of the specific models mentioned above. 
Currently, such models for lepton-lepton and lepton-quark contact interactions 
are constrained along this line, for the different flavours, by experiments 
on lepton and hadron production in $e^+e^-$ annihilation, unpolarized and 
polarized deep inelastic scattering, production of Drell-Yan lepton pairs, 
and atomic parity violation, see, e.g., 
Refs.~\cite{Barger:1998nf} and \citer{Abbaneo:2000nr,Barger:2000gv}. 
Global analyses are often performed for individual lepton-quark CI's, by 
combining data from various experiments and processes relevant to the same 
kind of coupling. The resulting bounds on $\Lambda$'s are quite sensitive 
to the particular one-parameter scenario considered in the analysis of 
the data and, without entering into details, typically range between 5 and 
12 TeV. 
\par    
On the other hand, in principle, constraints obtained by simultaneously 
including couplings of different chiralities might become considerably 
weaker, as noted above and also in Ref.~\cite{Zarnecki:1999yv}. Therefore, it 
should be highly desirable to apply a more general (and model-independent) 
approach to the analysis of experimental data, that simultaneously 
includes all terms of Eq.~(\ref{lagra}) as independent free parameters, 
and can also allow the derivation of separate constraints (or exclusion 
regions) on the values of the coupling constants.
\par 
To this aim, in the case of the process (\ref{proc}) at the LC 
considered here, 
a possibility is offered by initial beam polarization, that 
enables us to extract from the data the individual helicity cross sections 
through the definition of particular, and optimal, polarized integrated cross
sections and, consequently, to disentangle the constraints on 
the corresponding CI constants 
\citer{Pankov:1998ad,Babich:2001ik}. 
In this article, we wish to to present a model-independent analysis of the CI 
that complements that of Refs.~\citer{Pankov:1998ad,Babich:2001ik}, and is 
based instead on the measurements of more `conventional' observables 
(but still assuming polarized electron and positron beams) 
such as the total cross section, the forward-backward asymmetry $A_{\rm FB}$, 
the left-right asymmetry $A_{\rm LR}$, and left-right forward-backward  
asymmetry $A_{\rm LR,FB}$. A particularly delicate point in the derivation of 
bounds is, in this case, the correlation among uncertainties on  
the different observables, that will be taken into account {\it via} 
the method of the covariance matrix. 

The remainder of the paper is organized as follows.
In Sect.~2 we discuss the observables being considered,
which are the conventional observables, as opposed to the
more delicate optimal observables considered elsewhere
\citer{Babich:2000kp,Babich:2001ik}.
In Sect.~3 we present an error analysis together with the numerical
input and the resulting bounds on contact-interaction parameters.

\section{Observables}
For the process Eq.~(\ref{proc}), limiting ourselves to the cases 
$f\neq e, t$, we can neglect fermion masses with respect to $\sqrt s$, 
and express the amplitude in the Born approximation including the 
$\gamma$ and $Z$ $s$-channel exchanges plus the contact-interaction term of 
Eq.~(\ref{lagra}). With $P_e$ and $P_{\bar e}$ the longitudinal polarizations 
of the electron and positron beams, respectively, and $\theta$ the angle 
between the incoming electron and the outgoing fermion in the c.m.\ frame, 
the differential cross section can be expressed as \cite{Schrempp:1988zy}: 
\begin{equation}
\frac{\dd\sigma}{\dd\cos\theta}
=\frac{3}{8}
\left[(1+\cos\theta)^2 {\sigma}_+
+(1-\cos\theta)^2 {\sigma}_-\right].
\label{cross}
\end{equation}
In terms of the helicity cross sections $\sigma_{\alpha\beta}$ (with
$\alpha,\beta={\rm L,R}$), directly related to the individual CI 
couplings $\epsilon_{\alpha\beta}$: 
\begin{eqnarray}
{\sigma}_{+}&=&\frac{1}{4}\,
\left[(1-P_e)(1+P_{\bar{e}})\,\sigma_{\rm LL}
+(1+P_e)(1- P_{\bar{e}})\,\sigma_{\rm RR}\right]\nonumber \\
&=&\frac{D}{4}\,\left[(1-P_{\rm eff})\,\sigma_{\rm LL}
+(1+P_{\rm eff})\,\sigma_{\rm RR}\right], 
\label{s+} \\
{\sigma}_{-}&=&\frac{1}{4}\,
\left[(1-P_e)(1+ P_{\bar{e}})\,\sigma_{\rm LR}
+(1+P_e)(1-P_{\bar{e}})\,\sigma_{\rm RL}\right] \nonumber \\
&=&
\frac{D}{4}\,\left[(1-P_{\rm eff})\,\sigma_{\rm LR}
+(1+P_{\rm eff})\,\sigma_{\rm RL}\right], \label{s-}
\end{eqnarray}
where 
\begin{equation}
P_{\rm eff}=\frac{P_e-P_{\bar{e}}}{1-P_eP_{\bar{e}}} 
\label{pg}
\end{equation}
is the effective polarization \cite{Flottmann:1995ga}, 
$\vert P_{\rm eff}\vert\leq 1$, and 
$D=1-P_eP_{\bar{e}}$. For unpolarized positrons 
$P_{\rm eff}\rightarrow P_e$ and $D\rightarrow 1$,
but with $P_{\bar{e}}\ne0$, $\vert P_{\rm eff}\vert$
can be larger than $|P_e|$.
Moreover, in Eqs.~(\ref{s+}) and (\ref{s-}):  
\begin{equation}
\sigma_{\alpha\beta}=N_C\sigma_{\rm pt}
\vert{\cal M}_{\alpha\beta}\vert^2,
\label{helcross}
\end{equation}
where $N_C\approx 3(1+\alpha_s/\pi)$ for quarks and $N_C=1$ for leptons, 
respectively, and $\sigma_{\rm pt}\equiv\sigma(e^+e^-\to\gamma^\ast\to l^+l^-)
=(4\pi\alpha^2)/(3s)$.
The helicity amplitudes ${\cal M}_{\alpha\beta}$ can be written as
\begin{equation}
{\cal M}_{\alpha\beta}=Q_e Q_f+g_\alpha^e\,g_\beta^f\,\chi_Z+
\frac{s}{\alpha}\,\epsilon_{\alpha\beta}
\label{amplit}
\end{equation}
where $\chi_Z=s/(s-M^2_Z+iM_Z\Gamma_Z)$ represents the $Z$ propagator,
$g_{\rm L}^f=(I_{3L}^f-Q_f s_W^2)/s_W c_W$ and 
$g_{\rm R}^f=-Q_f s_W^2/s_W c_W$ 
are the SM left- and right-handed fermion couplings of the $Z$
with $s_W^2=1-c_W^2\equiv \sin^2\theta_W$ 
and $Q_f$ the fermion electric charge.
\par
We now define, with $\epsilon$ the experimental efficiency for
detecting the final state under consideration, the four, directly measurable, 
integrated event rates:
\begin{equation}
N_{\rm L,F},\quad  N_{\rm R,F},\quad N_{\rm L,B},\quad N_{\rm R,B},
\label{obsn}
\end{equation}
where ($\alpha={\rm L,R}$)
\begin{equation}
N_{\alpha,{\rm F}}=\frac{1}{2}\Lumint\,\epsilon
\int_{0}^{1}(\dd\sigma_\alpha/\dd\cos\theta)\dd\cos\theta,
\label{nf}
\end{equation}
\begin{equation}
N_{\alpha,{\rm B}}=\frac{1}{2}\Lumint\,\epsilon
\int_{-1}^{0}(\dd\sigma_\alpha/\dd\cos\theta)\dd\cos\theta,
\label{nb}
\end{equation}
and subscripts R and L correspond to two sets of beam polarizations,
$P_e=+P_1$, $P_{\bar e}=-P_2$ ($P_{1,2}>0$) and $P_e=-P_1$,
$P_{\bar e}=+P_2$, respectively, or, alternatively, $P_{\rm eff}=\pm P$
with $D$ fixed. In Eqs.~(\ref{nf}) and (\ref{nb}), $\Lumint$ is the
time-integrated luminosity, we assume it to be equally distributed over 
the two combinations of beam polarizations, L and R. 
\par 
The set of `conventional' observables we consider here for the discussion of 
bounds on the CI parameters are the unpolarized cross section:
\begin{equation}
\label{unpol}
\sigma_{\rm unpol}
=\frac{1}{4}\left[\sigma_{\rm LL}+\sigma_{\rm LR}
+\sigma_{\rm RR}+\sigma_{\rm RL}\right]; 
\end{equation}
the (unpolarized) forward-backward asymmetry:
\begin{equation}
A_{\rm FB}=
\frac{3}{4}\,
\frac{\sigma_{\rm LL}-\sigma_{\rm LR}+\sigma_{\rm RR}-\sigma_{\rm RL}}
     {\sigma_{\rm LL}+\sigma_{\rm LR}+\sigma_{\rm RR}+\sigma_{\rm RL}}; 
\label{afbth}
\end{equation}    
the left-right and the left-right forward-backward asymmetries
(which both require polarization), that can be written as, respectively:
\begin{equation}
A_{\rm LR}=
\frac{\sigma_{\rm LL}+\sigma_{\rm LR}-\sigma_{\rm RR}-\sigma_{\rm RL}}
     {\sigma_{\rm LL}+\sigma_{\rm LR}+\sigma_{\rm RR}+\sigma_{\rm RL}},
\label{alrth}
\end{equation}
and 
\begin{equation}
A_{\rm LR,FB}=
\frac{3}{4}\,
\frac{\sigma_{\rm LL}-\sigma_{\rm RR}+\sigma_{\rm RL}-\sigma_{\rm LR}}
{\sigma_{\rm LL}+\sigma_{\rm RR}+\sigma_{\rm RL}+\sigma_{\rm LR}}.
\label{alrfbth}
\end{equation}
Using Eqs.~(\ref{helcross}) and (\ref{amplit}), one can easily express 
the deviations of these observables from the SM predictions in terms of the 
SM couplings and the CI couplings $\epsilon_{\alpha\beta}$ of 
Eq.~(\ref{lagra}).  
\par
The above observables are connected to the measured ones through the 
integrated event rates $N_{\rm L,R;F,B}$, see Eq.~(\ref{obsn}), 
as follows: 
\begin{equation}
D\,\sigma_{\rm unpol}=\frac{N_{\rm tot }^{\rm exp}}{\Lumint\,\epsilon},
\label{ntotth}
\end{equation}
where 
\begin{equation}
N_{\rm tot}^{\rm exp}=N_{\rm L,F}+N_{\rm R,F}+N_{\rm L,B}+N_{\rm R,B}
\label{ntot}
\end{equation}
is the total number of events observed with polarized beams
(for the four measurements).
Eq.~(\ref{ntotth}) expresses the well-known fact that, when both 
the electron and positron beams are  polarized, the total annihilation 
cross section into fermion-antifermion pairs will be increased by 
the factor $D$, with $1\leq D\leq 2$.
\par
For the experimental forward-backward asymmetry:
\begin{equation}
A_{\rm FB}=A_{\rm FB}^{\rm exp}\equiv
\frac{N_{\rm L,F}+N_{\rm R,F}-N_{\rm L,B}-N_{\rm R,B}}
{N_{\rm L,F}+N_{\rm R,F}+N_{\rm L,B}+N_{\rm R,B}}, 
\label{afbexp}
\end{equation}
Finally, for the experimental left-right and left-right forward-backward 
asymmetries the relations are  
\begin{equation}
P_{\rm eff}A_{\rm LR}=A_{\rm LR}^{\rm exp}\equiv
\frac{N_{\rm L,F}+N_{\rm L,B}-N_{\rm R,F}-N_{\rm R,B}}
           {N_{\rm L,F}+N_{\rm L,B}+N_{\rm R,F}+N_{\rm R,B}},
\label{alrexp}
\end{equation}
and 
\begin{equation}
P_{\rm eff}A_{\rm LR,FB}=A_{\rm LR,FB}^{\rm exp}\equiv
\frac{(N_{\rm L,F}-N_{\rm R,F})-(N_{\rm L,B}-N_{\rm R,B})}
            {N_{\rm L,F}+N_{\rm R,F} + N_{\rm L,B}+N_{\rm R,B}}. 
\label{alrfbexp}
\end{equation}
\par  
In the following analysis, cross sections will be evaluated including 
initial- and final-state radiation by means of the program 
ZFITTER \cite{zfitter}, which has to be used along with ZEFIT, adapted to 
the present discussion, with $m_{\rm top}=175$~GeV and
$m_H=120$~GeV. One-loop SM electroweak corrections are accounted for by 
improved Born amplitudes \cite{Consoli:1989pc,Altarelli:1990dt}, 
such that the forms of the 
previous formulae remain the same. Concerning initial-state radiation, a cut 
on the energy of the emitted photon $\Delta=E_\gamma/E_{\rm beam}=0.9$ is 
applied for $\sqrt s=0.5\ {\rm TeV}$ in order to avoid the radiative return 
to the $Z$ peak, and increase the signal originating from the contact 
interaction contribution \cite{Djouadi:1992sx}.                      
\par                 
As numerical inputs, we shall assume the commonly used
reference values of the identification efficiencies 
\cite{Damerell}:
$\epsilon=95\%$ for $l^+l^-$; $\epsilon=60\%$ for $b\bar{b}$; 
$\epsilon=35\%$ for $c\bar{c}$. Concerning the statistical 
uncertainty, to study the relative roles of statistical and systematic 
uncertainties we shall vary $\Lumint$ from $50$ to
$500\ \mbox{fb}^{-1}$ (half for each polarization orientation) with 
uncertainty $\delta\Lumint/\Lumint=0.5\%$, and a fiducial experimental 
angular range $|\cos\theta|\le 0.99$. Also, regarding electron and positron 
degrees of polarization, we shall consider the values: $\vert P_e\vert=0.8$; 
$\vert P_{\bar e}\vert=0.0,\ 0.4\ {\rm and}\ 0.6$, with  
$\delta P_e/P_e=\delta P_{\bar e}/P_{\bar e}=0.5\ \%$.
\section{Numerical analysis and constraints on CI couplings}
The current bounds on $\Lambda_{\alpha\beta}$ cited in Sect.~1, 
of the order of several TeV, are such that for the LC c.m.\ energy 
$\sqrt{s}=0.5$~TeV the characteristic suppression factor $s/\Lambda^2$ 
in Eq.~(\ref{amplit}) is rather strong. Accordingly, we can safely assume 
a linear dependence of the cross sections on the parameters 
$\epsilon_{\alpha\beta}$. In this regard, indirect manifestations of 
the CI interaction (\ref{lagra}) can be looked for, {\it via} deviations 
of the measured observables from the SM predictions, caused by the 
new interaction. 
The reach on the CI couplings, and the corresponding constraints on their 
allowed values in the case of no effect observed, can be estimated 
by comparing the expression of the mentioned deviations with the expected 
experimental (statistical and systematic) uncertainties. 
\par      
To this purpose, assuming the data to be well described by the SM 
($\epsilon_{\alpha\beta}=0$) predictions, i.e., that no deviation is 
observed within the foreseen experimental uncertainty, and in the linear 
approximation in $\epsilon_{\alpha\beta}$ of the observables 
(\ref{unpol})--(\ref{alrfbth}), we apply the method based on the covariance 
matrix:
\begin{eqnarray}
\label{covarv}
V_{kl}&=&\langle({\cal O}_k-\bar{\cal O}_k)
                ({\cal O}_l-\bar{\cal O}_l)\rangle\nonumber \\
&=&
\sum_{i=1}^{4}\left(\delta N_i\right)^2  
\left(\frac{\partial{\cal O}_k}{\partial N_i}\right)
\left(\frac{\partial{\cal O}_l}{\partial N_i}\right)
+\left(\delta \Lumint \right)^2  
\left(\frac{\partial{\cal O}_k}{\partial\Lumint}\right)
\left(\frac{\partial{\cal O}_l}{\partial\Lumint}\right)
\nonumber \\
&+&
\left(\delta P_e\right)^2  
\left(\frac{\partial{\cal O}_k}{\partial P_e}\right)
\left(\frac{\partial{\cal O}_l}{\partial P_e}\right)+
\left(\delta P_{\bar e}\right)^2  
\left(\frac{\partial{\cal O}_k}{\partial P_{\bar e}}\right)
\left(\frac{\partial{\cal O}_l}{\partial P_{\bar e}}\right).
\end{eqnarray}
Here, the $N_i$ are given by Eq.~(\ref{obsn}), so that the statistical 
error appearing on the right-hand-side is given by
\begin{equation}
\delta N_i=\sqrt{N_i}, \label{deltani} \end{equation}
and the ${\cal O}_l=(\sigma_{\rm unpol}$, $A_{\rm FB}$, $A_{\rm LR}$, 
$A_{\rm LR,FB})$ are the four observables. The second, third and fourth 
terms of the right-hand-side of Eq.~(\ref{covarv}) represent the systematic 
errors on the integrated luminosity $\Lumint$, polarizations $ P_e$ and 
$P_{\bar e}$, respectively, for which we assume the numerical values reported
in the previous Section. From the explicit expression of the matrix 
elements $V_{kl}$, one can easily notice that, apart from 
$\sigma_{\rm unpol}$ and $A_{\rm FB}$ that are uncorrelated ($V_{12}=0$), 
all other pairs of observables show a correlation. Indeed, the non-zero 
diagonal entries are given by:
\begin{eqnarray}
\label{Eq:V-diag}
&&V_{11}= \frac{\sigma_{\rm unpol}^2}{N_{\rm tot}^{\rm exp}} 
+ \sigma_{\rm unpol}^2 \, 
\left[ \frac{P^2_e P^2_{\bar e}}{D^2}(\epsilon^2_e +\epsilon^2_{\bar e}) 
+ \epsilon^2_{\cal L} \right]; \qquad 
V_{22}=\frac{1-A^2_{\rm FB}}{N_{\rm tot}^{\rm exp}}, \nonumber\\
&& 
V_{33}=\frac{1-A^2_{\rm LR} P^2_{\rm eff}}{P^2_{\rm eff} N_{\rm tot}^{\rm exp}}
+ A^2_{\rm LR} \, \Delta^2_2;\qquad 
V_{44}=\frac{1-A^2_{\rm LR,FB} P^2_{\rm eff}}
{P^2_{\rm eff} N_{\rm tot}^{\rm exp}} + A^2_{\rm LR,FB}\, \Delta^2_2,
\label{diagv}
\end{eqnarray}
and, for the non-diagonal ones we have: 
\begin{eqnarray} 
&&V_{13}= \sigma_{\rm unpol} \, A_{\rm LR} \, \Delta^2_1; \qquad 
V_{14}= \sigma_{\rm unpol}\,
A_{\rm LR, FB} \, \Delta^2_1; \qquad 
V_{23}=\frac{A_{\rm LR,FB}-A_{\rm FB} A_{\rm LR}}{N_{\rm tot}^{\exp}}, 
\nonumber \\
&&V_{24}=\frac{A_{\rm LR}-A_{\rm FB} A_{\rm LR,FB}}{N_{\rm tot}^{\rm exp}}; 
\quad  
V_{34}=\frac{A_{\rm FB}-A_{\rm LR} A_{\rm LR,FB} 
P^2_{\rm eff}}{P^2_{\rm eff} N_{\rm tot}^{\rm exp}} + A_{\rm LR}\, 
A_{\rm LR,FB}\, \Delta^2_2. 
\label{nondiagv}
\end{eqnarray}
Here: 
\begin{eqnarray}
&&\Delta^2_1=\frac{P_e \, P_{\bar e}}{P_{\rm eff} D^3} 
\left[-(1-P^2_{\bar e})\, P_e \, \epsilon^2_e + (1-P^2_e)\,
P_{\bar e} \, \epsilon^2_{\bar e} \right],  \nonumber \\
&&\Delta^2_2=\frac{(1-P^2_{\bar e})^2 \, P^2_e \, \epsilon^2_e
+ (1-P^2_e)^2 \, P^2_{\bar e} \, \epsilon^2_{\bar e}}{P^2_{\rm eff}D^4},
\end{eqnarray}
and $\epsilon_e=\delta P_e/P_e$, 
$\epsilon_{\bar e}=\delta P_{\bar e}/P_{\bar e}$ 
and $\epsilon_{\cal L}=\delta {\cal L}_{\rm int}/{\cal L}_{\rm int}$ 
are the relative systematic uncertainties. 
\par 
One can notice, from Eq.~(\ref{Eq:V-diag}), that systematic 
uncertainties in $\sigma_{\rm unpol}$ are induced by 
$\epsilon_e$, $\epsilon_{\bar e}$ {\it and} $\epsilon_{\cal L}$, while those 
in $A_{\rm LR}$ and $A_{\rm LR,FB}$ arise from  
$\epsilon_e$ and $\epsilon_{\bar e}$ only, and {\it not} from 
$\epsilon_{\cal L}$.  
Finally, $A_{\rm FB}$ is free from such systematic uncertainties. 

Defining the inverse covariance matrix $W^{-1}$ as 
\begin{equation}
\left(W^{-1}\right)_{ij}=\sum_{k,l=1}^4 \left(V^{-1}\right)_{kl}
\left(\frac{\partial{\cal O}_k}{\partial\epsilon_i}\right)
\left(\frac{\partial{\cal O}_l}{\partial\epsilon_j}\right), 
\end{equation}
with 
$\epsilon_i=(\epsilon_{LL},\ \epsilon_{LR},\ \epsilon_{RL},\ \epsilon_{RR})$, 
model-independent allowed domains in the four-dimensional CI parameter space
to 95\% confidence level are obtained from the error contours determined by 
the quadratic form in $\epsilon_{\alpha\beta}$ \cite{eadie,Cuypers:1996it}:
\begin{equation}
\label{chi2}
\left(\epsilon_{LL}\ \epsilon_{LR}\ \epsilon_{RL}\ \epsilon_{RR}\right) 
W^{-1}
\left( \begin{array}{c}
\epsilon_{LL}\\
\epsilon_{LR}\\
\epsilon_{RL}\\
\epsilon_{RR}
\end{array}\right)=9.49.
\end{equation}
The value 9.49 on the right-hand side of Eq.~(\ref{chi2}) corresponds to 
a fit with four free parameters \cite{Groom:2000in,James:1975dr}. 

The quadratic form (\ref{chi2}) defines a four-dimensional ellipsoid in the 
$\left(\epsilon_{LL},\ \epsilon_{LR},\ \epsilon_{RL},\ \epsilon_{RR}\right)$
parameter space. 
The matrix $W$ has the property that the square roots of the individual   
diagonal matrix elements, $\sqrt{W_{ii}}$, determine the projection 
of the ellipsoid onto the corresponding $i$-parameter axis in the 
four-dimensional space, and has the meaning of the bound at one~$\sigma$ on 
that parameter regardless of the values assumed for the others. 
Conversely, $1/\sqrt{\left(W^{-1}\right)_{ii}}$ determines the value
of the intersection of the ellipsoid with the corresponding $i$-parameter 
axis, and represents the one-$\sigma$ bound on that parameter assuming 
all the others to be exactly known and equal to zero. Accordingly, 
with the R.H.S.\ of Eq.~(\ref{chi2}), the 
ellipsoidal surface constrains, at the 95\% C.L. and model-independently, 
the range of values of the CI couplings $\epsilon_{\alpha\beta}$ allowed 
by the foreseen experimental uncertainties.    
\par 
For the chosen input values for integrated luminosity, initial beam
polarization, and corresponding systematic uncertainties, such 
model-independent limits are listed as lower bounds on the mass scales 
$\Lambda_{\alpha\beta}$ in Table~1.
\par 
All the numerical results exhibited in Table~1 can be represented graphically.
In Figs.~1--3 we show the planar ellipses that are obtained by projecting
onto the six planes 
($\epsilon_{LL},\epsilon_{LR}$), ($\epsilon_{LL},\epsilon_{RL}$), 
($\epsilon_{LL},\epsilon_{RR}$), ($\epsilon_{RR},\epsilon_{LR}$), 
($\epsilon_{RR},\epsilon_{RL}$), ($\epsilon_{LR},\epsilon_{RL}$) 
the 95\% C.L. allowed four-dimensional ellipsoid resulting from 
Eq.~(\ref{chi2}). In these figures, the inner and outer ellipses correspond 
to positron polarizations $\vert P_{\bar e}\vert=0.6$ and 
$\vert P_{\bar e}\vert=0.0$, respectively. 

\begin{table}[ht]
\centering
\caption{
Reach in $\Lambda_{\alpha\beta}$ at 95\% C.L., for the
model-independent analysis performed in terms of conventional observables,
for $e^+e^-\to\mu^+\mu^-, b \bar b$ and
$c \bar c$ at $E_{\rm c.m.}=0.5$~TeV, $\Lumint=50\,\mbox{fb}^{-1}$ and
$500\,\mbox{fb}^{-1}$, $\vert P_{e}\vert=0.8$ and 
$\vert P_{\bar e}\vert=0.0; 0.4; 0.6$.
}
\medskip
{\renewcommand{\arraystretch}{1.2}
\begin{tabular}{|c|c|c|c|c|c|}
\hline
$f\bar f$ & ${\Lumint}$ & $\Lambda_{\rm LL}$ & $\Lambda_{\rm RR}$ &
$\Lambda_{\rm LR}$ & $\Lambda_{\rm RL}$ \\
&$\mbox{fb}^{-1}$& TeV & TeV &TeV  & TeV \\
\hline
\cline{2-6}
 & 50 &$31.2; 33.9; 34.8$&$30.9; 33.9; 35.0$
 &$26.5; 29.4; 30.5$&$26.9; 29.9; 31.1$
\\
\cline{2-6}
$\mu^+\mu^-$
 & 500 &$46.1; 46.9; 46.8$&$47.0; 48.7; 48.7$
 &$45.8; 49.9; 51.3$&$46.5; 50.7; 52.0$ 
\\ \hline
\cline{2-6}
 & 50 &$37.0; 39.8; 40.8$&$33.0; 39.1; 41.4$
 &$25.2; 27.7; 28.7$&$31.8; 36.2; 37.9$ 
\\
\cline{2-6}
$b\bar b$
 & 500 &$53.1; 54.0; 54.0$&$50.5; 63.9; 68.3$
 &$44.5; 48.8; 50.4$&$56.0; 63.3; 66.0$ 
\\ \hline
\cline{2-6}
 & 50 &$32.1; 34.7; 35.7$&$31.4; 35.4; 36.8$
 &$26.2; 28.8; 29.8$&$21.6; 24.5; 25.7$ 
\\
\cline{2-6}
$c\bar c$
& 500 &$47.6; 48.4; 48.4$&$50.0; 54.9; 55.6$
&$45.7; 49.7; 51.2$&$38.1; 43.4; 45.5$
\\ \hline
\end{tabular}
} 
\label{tab:table-1}
\end{table}

To appreciate the significant role of initial beam polarization we 
should consider that, in the unpolarized case the only available 
observables would be 
$\sigma_{\rm unpol}\propto (\sigma_{\rm LL}+\sigma_{\rm RR})
+(\sigma_{\rm LR}+\sigma_{\rm RL})$ and 
$\sigma_{\rm unpol}\cdot A_{\rm FB}\propto (\sigma_{\rm LL}+\sigma_{\rm RR})
-(\sigma_{\rm LR}+\sigma_{\rm RL})$, see Eqs.~(\ref{unpol}) and 
(\ref{afbth}). Therefore, by themselves, the pair of experimental 
observables $\sigma_{\rm unpol}$ and $A_{\rm FB}$ are not able to limit 
separately the CI couplings within finite ranges, but could only provide 
a constraint among the {\it linear combinations} of parameters  
$(\epsilon_{\rm LL}+\epsilon_{\rm RR})$ and 
$(\epsilon_{\rm LR}+\epsilon_{\rm RL})$. In some planes, specifically in 
the ($\epsilon_{\rm LL},\epsilon_{\rm RR})$ and 
$(\epsilon_{\rm LR},\epsilon_{\rm RL})$ planes, this constraint has the form 
of (unlimited) bands of allowed values, or correlations, such as those 
limited by the straight lines in Figs.~1--3. 
With initial beam polarization, two more physical observables become 
available, {\it i.e.}, $A_{\rm LR}$ and $A_{\rm LR,FB}$, and this enables us 
to close the bands into the ellipses in Figs.~1--3. 
The allowed bounds obtained from the observables $\sigma_{\rm unpol}$ and
$A_{\rm FB}$ are not affected by electron polarization (for unpolarized
positrons). Therefore, the bounds in the form of straight lines 
are tangential to the outer ellipses referring to $P_{\bar e}=0$, and in this
case the role of $P_e\ne 0$ is just to close the corresponding band to a 
finite area.

\par 
While, in order to achieve this result, electron polarization is 
obviously essential, the results in Table~1 show that, at least in the 
kind of analysis presented here, the increase in the sensitivity to 
the parameters $\epsilon_{\alpha\beta}$ due to the additional availability 
of positron polarization is appreciable in most cases, but very modest or 
even absent in some, depending on the final $f{\bar f}$ channel 
and on the luminosity. This situation occurs in the cases, limited to 
the diagonal LL and RR combinations, where, with our chosen numerical input,
the additional systematic uncertainty introduced by positron 
polarization competes with the statistical one. This is not the case for the 
non-diagonal combinations RL and LR, for which the statistical uncertainty 
remains the dominant one. 
\par 
Furthermore, from the orientations of the ellipsis axes in Figs.~1--3, one 
can observe that in most considered cases the correlation among the 
different pairs of CI couplings is not so pronounced, but 
becomes somewhat more important at the higher value of the integrated 
luminosity. Clearly, from the strictly quantitative point of view, these 
features relate to the assumed input values for the systematic 
uncertainties.
\par 
The crosses in Figs.~1--3 represent the constraints obtainable 
by taking only one non-zero parameter at a time, instead of all four
simultaneously non-zero, and independent, as in the analysis discussed here.
Similar to the inner and outer ellipses, the shorter and longer 
arms of the crosses refer to positron polarization 
$\vert P_{\bar e}\vert=0.6$ and 0.0, respectively.  Such `one-parameter' 
results are derived from a $\chi^2$ procedure applied to the combination 
of the four physical observables (\ref{unpol})--(\ref{alrfbth}), also 
taking the above-mentioned correlations among observables into 
account.\footnote{This procedure leads to results numerically consistent 
with those presented from essentially the same set of observables 
in Ref.~\cite{Riemann:2001bb}, 
if applied to the same experimental inputs used there.}
As regards the role of polarization in this kind of single-parameter 
analysis, one finds that, basically, positron polarization in addition to 
electron polarization has more impact and generally allows an appreciable 
improvement of the sensitivity to CI couplings. Finally, one can note 
from Figs.~1--3 that the `single-parameter' constraints on the individual 
CI parameters are numerically more stringent, as compared to the 
model-independent ones. Essentially, this is a reflection of the smaller 
value of the critical $\chi^2$, $\chi^2_{\rm crit}=3.84$ corresponding to 
95\% C.L. with a {\it one-parameter} fit, and also of a reduced role of 
correlations. 
\par 
On the other hand, as repeatedly emphasized in the Introduction, that
procedure allows to test individual models corresponding to the 
parameter (or the combination of parameters) assumed to be non-zero, and 
in this sense it is model-dependent, whereas the allowed areas represented 
by the ellipses in Figs.~1--3 are model-independent. Actually, in this 
regard, one could compare the sensitivities shown in Table 1 (and 
Figs.~1--3) with the model-independent constraints on the parameters 
$\epsilon_{\alpha\beta}$ previously obtained using a set of polarized cross 
sections integrated in `optimal' angular ranges as basic experimental 
observables \citer{Babich:2000kp,Babich:2001ik}. 
One finds that, with few exceptions 
relevant to specific cases in the $b{\bar b}$ and $c{\bar c}$ sectors, 
although defining a somewhat more involved procedure the latter observables 
produce more stringent constraints compared to the simpler (and more 
practical)  observables used here (mostly asymmetries). This is the combined 
effect of the relative numerical weights of statistical and systematical 
uncertainties input in the calculation, and of the optimization procedure 
introduced in \citer{Babich:2000kp,Babich:2001ik}. In principle, 
further optimization of 
the asymmetries $A_{\rm FB}$ and $A_{\rm LR,FB}$ may be possible.  

\medskip
\leftline{\bf Acknowledgements}
\par\noindent
This research has been supported by the Research Council of Norway,
by MURST (Italian Ministry of University, Scientific Research
and Technology) and by funds of the University of Trieste.
\goodbreak

\begin{figure}[htb]
\refstepcounter{figure}
\label{Fig:m}
\addtocounter{figure}{-1}
\begin{center}
\setlength{\unitlength}{1cm}
\begin{picture}(11.5,18)
\put(-1.0,8.0)
{\mbox{\epsfysize=10.0cm\epsffile{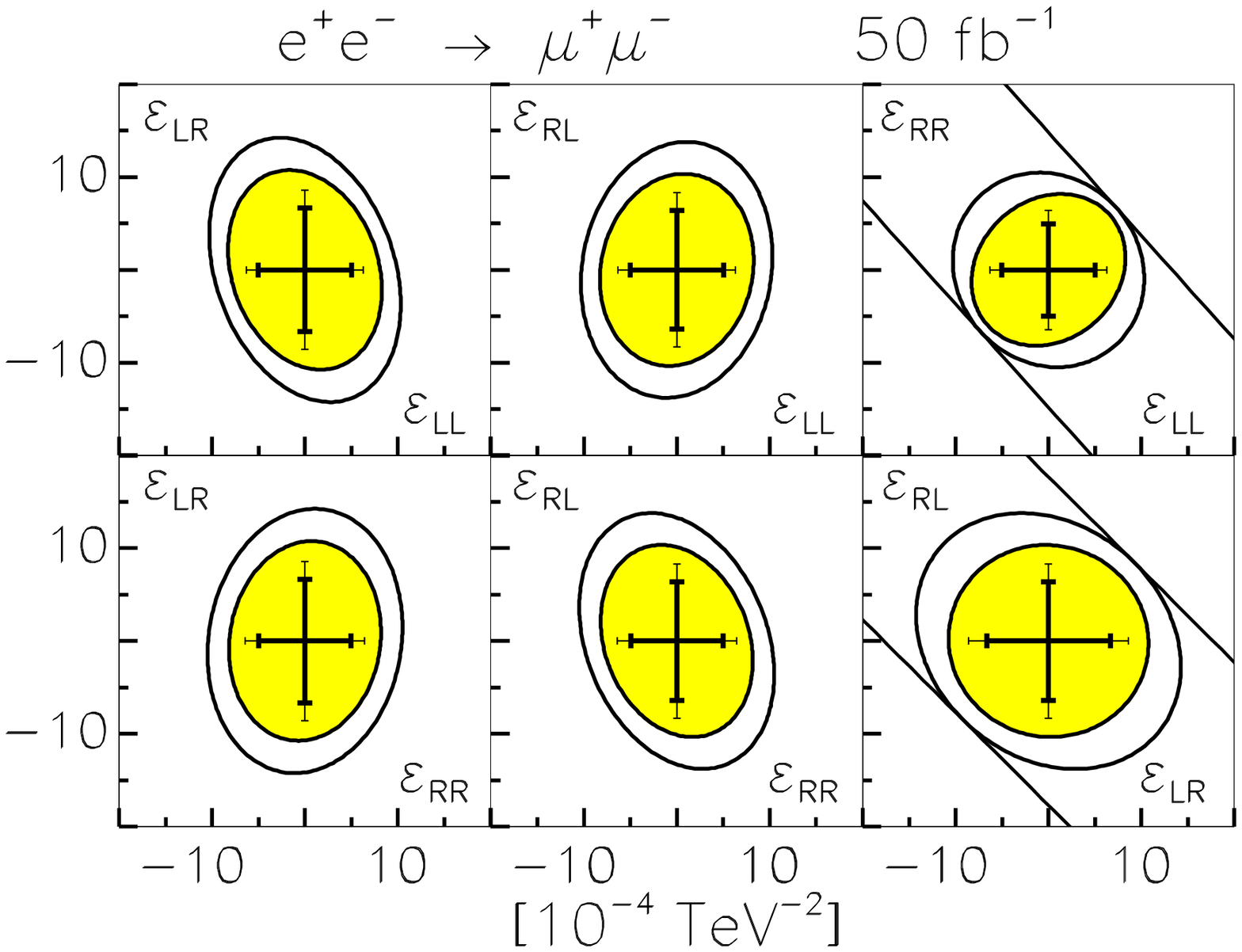}}}
\put(-1.0,-1.0)
{\mbox{\epsfysize=10.0cm\epsffile{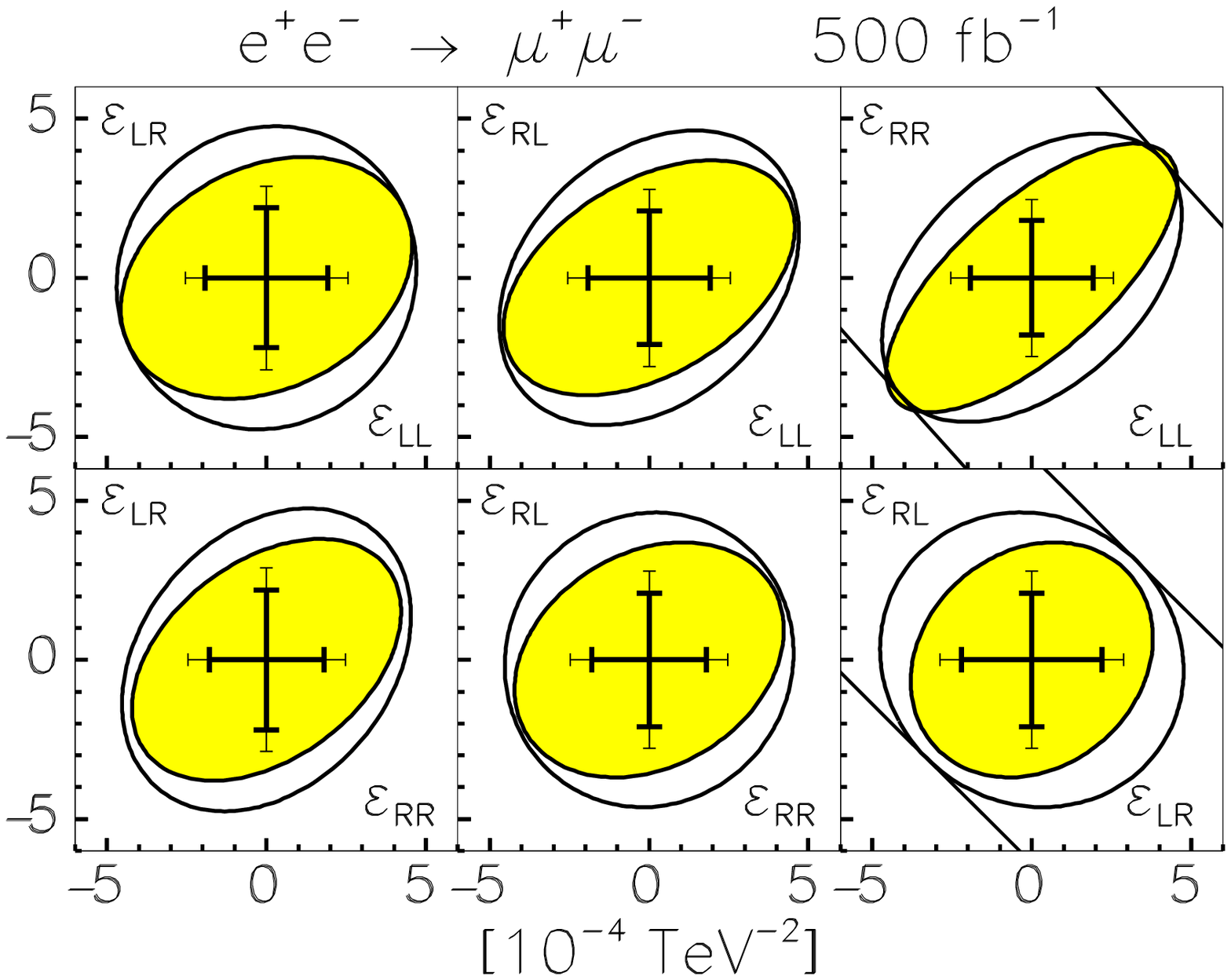}}}
\end{picture}
\caption{Two-dimensional projections of the 95\% C.L. allowed region 
(\ref{chi2}) for $e^+e^-\to\mu^+\mu^-$ at $\Lumint=50~\text{fb}^{-1}$ 
and $\Lumint=500~\text{fb}^{-1}$. Note that the scales are different.  
$\vert P_e\vert=0.8$, $\vert P_{\bar e}\vert=0.0$ (outer ellipse) and 
$\vert P_{\bar e}\vert=0.6$ (inner ellipse). The solid crosses represent  
the `one-parameter' bounds under the same conditions. 
}
\end{center}
\end{figure}

\begin{figure}[htb]
\refstepcounter{figure}
\label{Fig:b}
\addtocounter{figure}{-1}
\begin{center}
\setlength{\unitlength}{1cm}
\begin{picture}(11.5,18)
\put(-1.0,8.0)
{\mbox{\epsfysize=10.0cm\epsffile{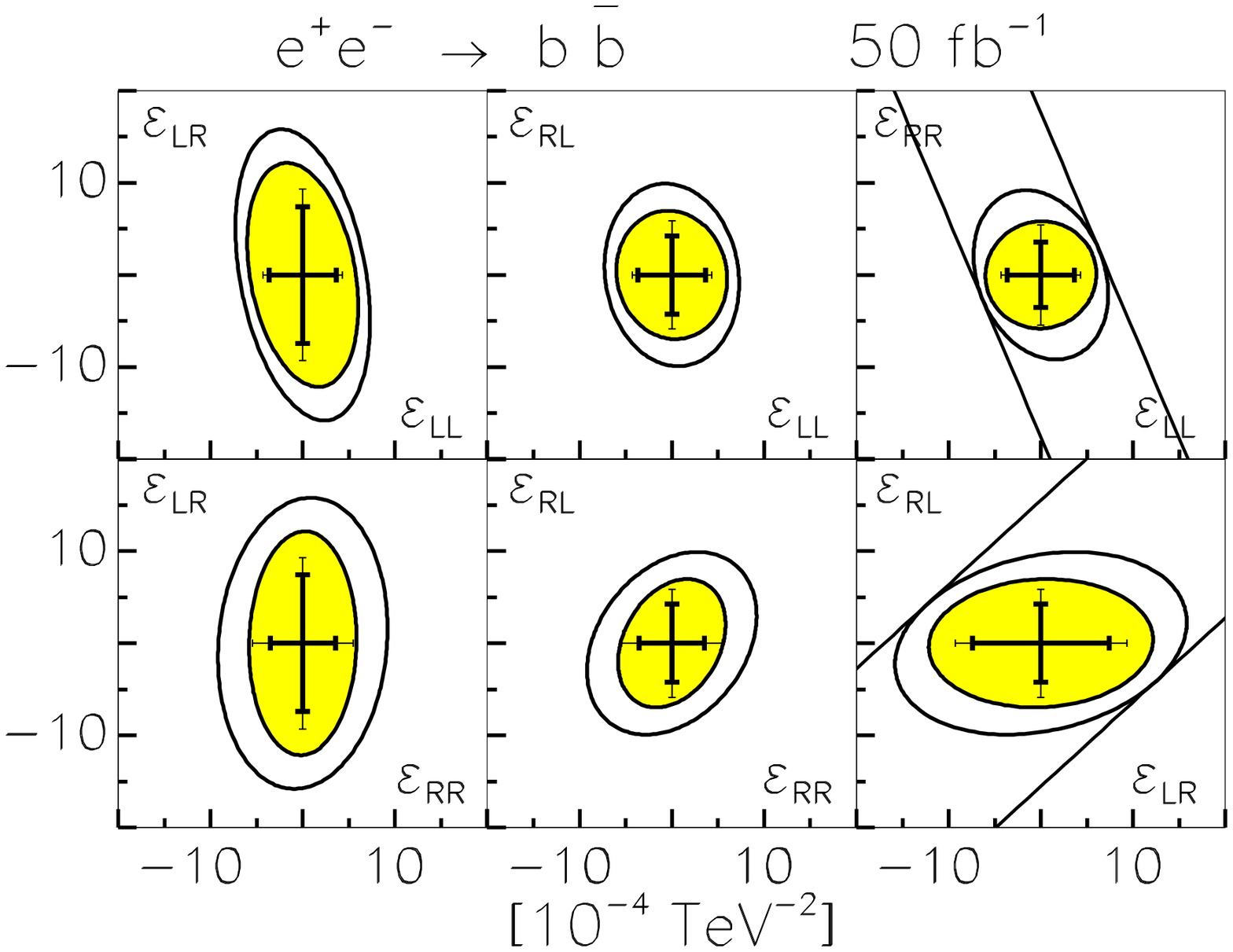}}}
\put(-1.0,-1.0)
{\mbox{\epsfysize=10.0cm\epsffile{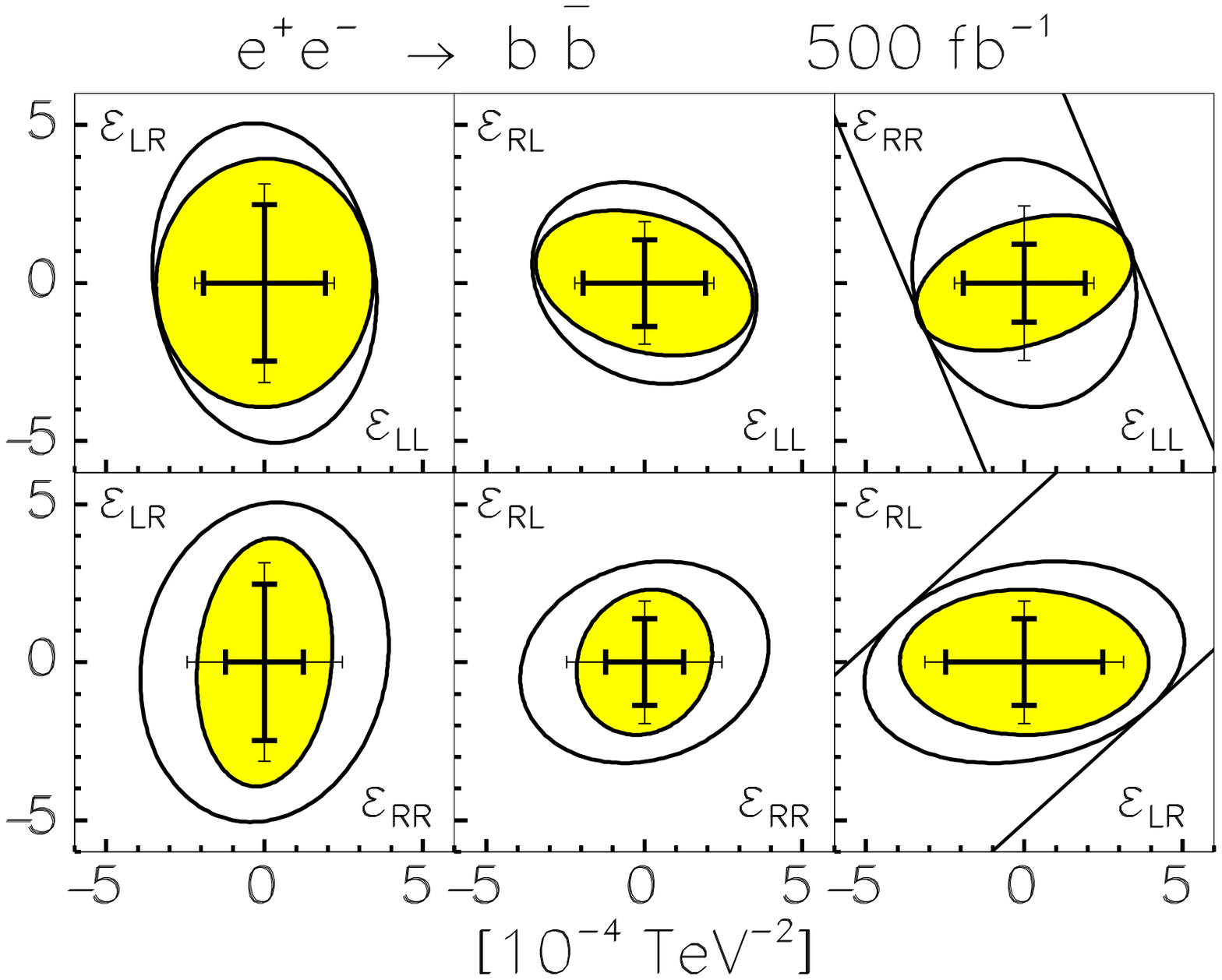}}}
\end{picture}
\caption{
Same as Fig.~1, for the process $e^+e^-\to b {\bar b}$.}
\end{center}
\end{figure}

\begin{figure}[htb]
\refstepcounter{figure}
\label{Fig:c}
\addtocounter{figure}{-1}
\begin{center}
\setlength{\unitlength}{1cm}
\begin{picture}(11.5,18)
\put(-1.0,8.0)
{\mbox{\epsfysize=10.0cm\epsffile{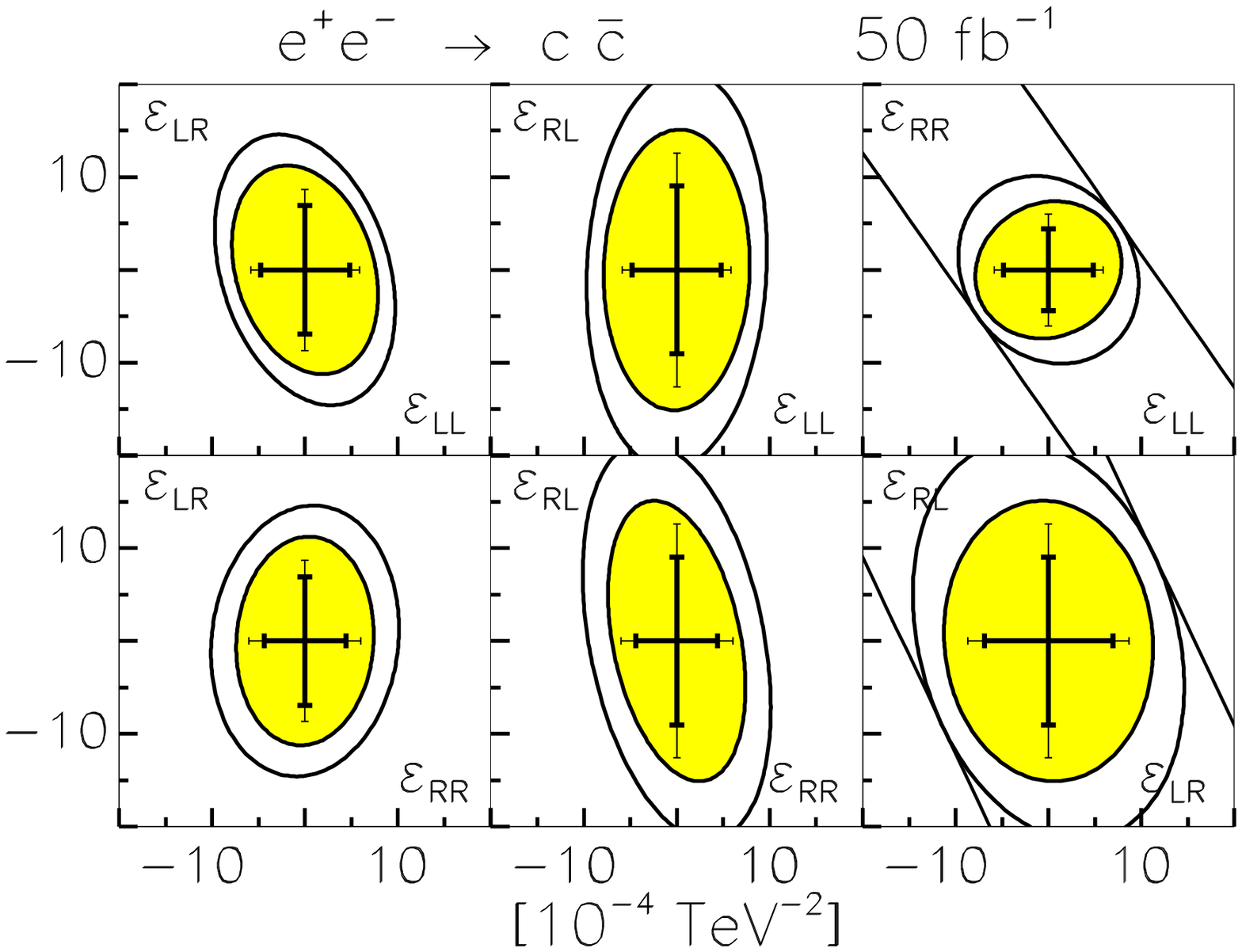}}}
\put(-1.0,-1.0)
{\mbox{\epsfysize=10.0cm\epsffile{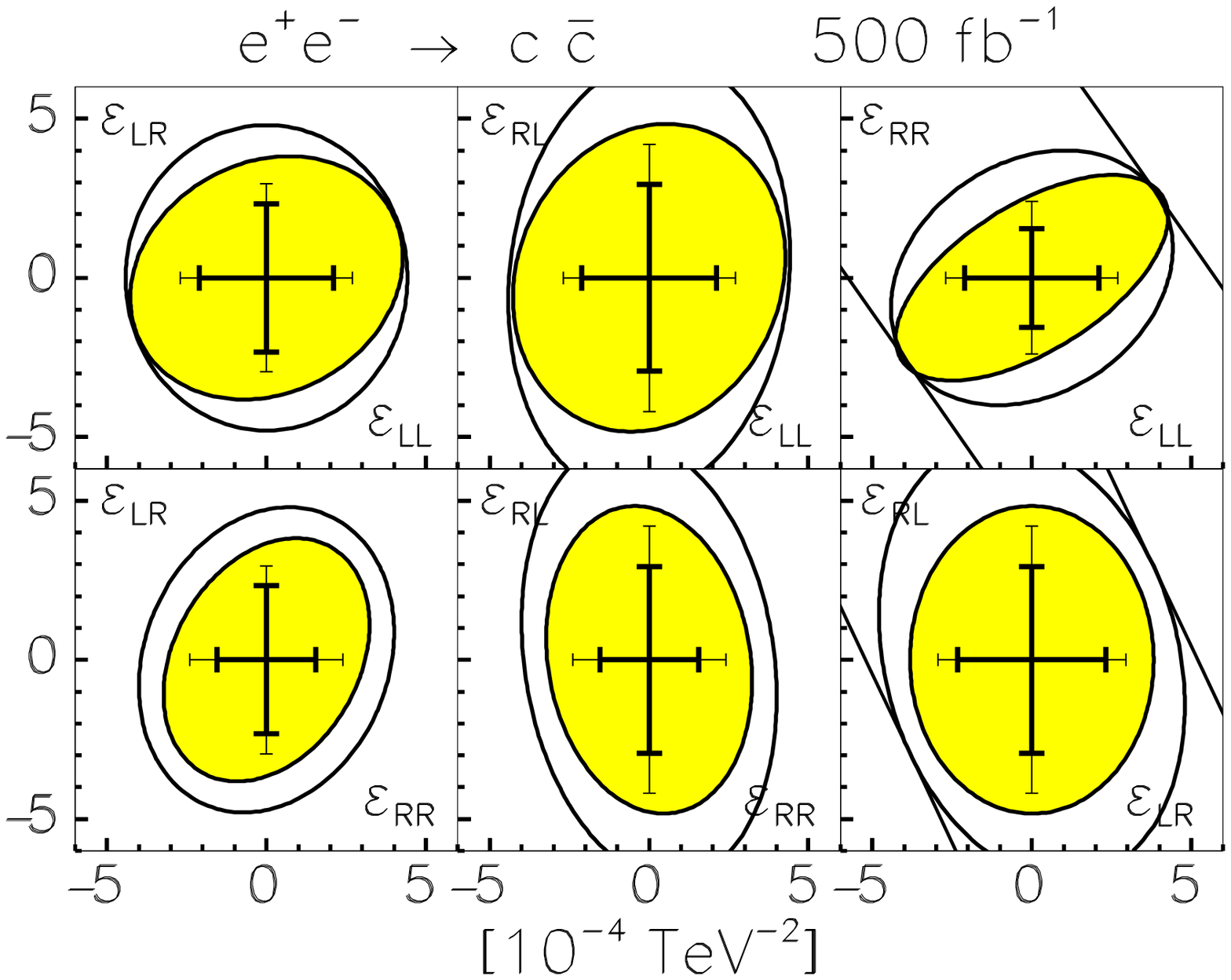}}}
\end{picture}
\caption{
Same as in Fig.~1, for the process $e^+e^-\to c {\bar c}$.}
\end{center}
\end{figure}



\begin{thebibliography}{99}

\bibitem{Barger:1998nf}
V.~Barger, K.~Cheung, K.~Hagiwara and D.~Zeppenfeld,
Phys.\ Rev.\ D {\bf 57} (1998) 391
[hep-ph/9707412]; \\
%
D.~Zeppenfeld and K.~Cheung,
hep-ph/9810277; \\
K.~Cheung, hep-ph/0106251.

\bibitem{Altarelli:1997ce}
G.~Altarelli, J.~Ellis, G.~F.~Giudice, S.~Lola and M.~L.~Mangano,
Nucl.\ Phys.\ B {\bf 506} (1997) 3
[hep-ph/9703276]; \\
%
R.~Casalbuoni, S.~De Curtis, D.~Dominici and R.~Gatto,
Phys.\ Lett.\ B {\bf 460} (1999) 135
[hep-ph/9905568].
 
\bibitem{Eichten:1983hw}
E.~Eichten, K.~Lane and M.~E.~Peskin,
Phys.\ Rev.\ Lett.\ {\bf 50} (1983) 811; \\
%
R.~R\"uckl,
Phys.\ Lett.\ B {\bf 129} (1983) 363.

\bibitem{Abbaneo:2000nr}
D.~Abbaneo {\it et al.}  [ALEPH Collaboration],
CERN-EP-2000-016.

\bibitem{Zarnecki:1999je}
A.~F.~Zarnecki,
Eur.\ Phys.\ J.\ C {\bf 11} (1999) 539
[hep-ph/9904334].

\bibitem{Zarnecki:1999yv}
A.~F.~Zarnecki,
Nucl.\ Phys.\ Proc.\ Suppl.\ {\bf 79} (1999) 158
[hep-ph/9905565].

\bibitem{Barger:2000gv}
V.~Barger and K.~Cheung,
Phys.\ Lett.\ B {\bf 480} (2000) 149
[hep-ph/0002259].

\bibitem{Pankov:1998ad}
A.~A.~Pankov and N.~Paver,
Phys.\ Lett.\ B {\bf 432} (1998) 159
[hep-ph/9805207].

\bibitem{Babich:2000kp}
A.~A.~Babich, P.~Osland, A.~A.~Pankov and N.~Paver,
Phys.\ Lett.\ B {\bf 476} (2000) 95
[hep-ph/9910403].

\bibitem{Babich:2000kx}
A.~A.~Babich, P.~Osland, A.~A.~Pankov and N.~Paver,
Phys.\ Lett.\ B {\bf 481} (2000) 263
[hep-ph/0003253].

\bibitem{Babich:2001ik}
A.~A.~Babich, P.~Osland, A.~A.~Pankov and N.~Paver,
LC Note LC-TH-2001-021 (2001), 
hep-ph/0101150.

\bibitem{Schrempp:1988zy}
B.~Schrempp, F.~Schrempp, N.~Wermes and D.~Zeppenfeld,
Nucl.\ Phys.\ B {\bf 296} (1988) 1.

\bibitem{Flottmann:1995ga}
K.~Flottmann,
DESY-95-064; 
%
K.~Fujii and T.~Omori,
KEK-PREPRINT-95-127.

\bibitem{zfitter}
S. Riemann, FORTRAN program ZEFIT Version 4.2; \\
D. Bardin et al., 
Comput.\ Phys.\ Commun.\  {\bf 133} (2001) 229
[hep-ph/9908433].

\bibitem{Consoli:1989pc}
M.~Consoli, W.~Hollik and F.~Jegerlehner,
CERN-TH-5527-89
{\it Presented at Workshop on Z Physics at LEP}.

\bibitem{Altarelli:1990dt}
G.~Altarelli, R.~Casalbuoni, D.~Dominici, F.~Feruglio and R.~Gatto,
Nucl.\ Phys.\ B {\bf 342} (1990) 15.

\bibitem{Djouadi:1992sx}
A.~Djouadi, A.~Leike, T.~Riemann, D.~Schaile and C.~Verzegnassi,
Z.\ Phys.\ C {\bf 56} (1992) 289.

\bibitem{Damerell}
C. J. S. Damerell, D.J. Jackson,
in {\it Proceedings of the 1996 DPF/DPB Summer
Study on New Directions for High Energy Physics} (Snowmass 96),
Edited by D.G. Cassel, L. Trindle Gennari, R.H. Siemann
(SLAC, 1997) p. 442.

\bibitem{eadie}
W.T. Eadie, D. Drijard, F.E. James, M. Roos, B. Sadoulet,
{\it Statistical methods in experimental physics}
(American Elsevier, 1971).

\bibitem{Cuypers:1996it}
F.~Cuypers and P.~Gambino,
Phys.\ Lett.\ B {\bf 388} (1996) 211
[hep-ph/9606391]; \\
F.~Cuypers,
hep-ph/9611336. 

\bibitem{Groom:2000in}
D.~E.~Groom {\it et al.}  [Particle Data Group Collaboration],
Eur.\ Phys.\ J.\ C {\bf 15} (2000) 1.

\bibitem{James:1975dr}
F.~James and M.~Roos,
Comput.\ Phys.\ Commun.\ {\bf 10} (1975) 343.

\bibitem{Riemann:2001bb}
S.~Riemann,
LC Note LC-TH-2001-007 (2001).

\end{thebibliography}
\end{document}